\documentclass[article,preprint,amsmath,amssymb,amsfonts,superscriptaddress]{revtex4-2}
\usepackage{times,graphicx,caption,epstopdf,dcolumn,bm,color,braket,multirow,xfrac,etoolbox}

\usepackage[colorlinks=true, linkcolor=blue, citecolor=blue, urlcolor=blue]{hyperref} %본문 인용 숫자를 클릭하면 레퍼런스 또는 그림으로 이동함
\usepackage[T1]{fontenc} %유럽 언어에서 발생할 수 있는 특수 문자나 악센트 문자를 적절하게 표시하고 처리
\usepackage{amsmath} %복잡한 수학식을 표현할 때 유용한 환경을 제공
\usepackage{amssymb}
\usepackage{xr-hyper} % for SI
\renewcommand{\figurename}{FIG.}
\renewcommand{\thefigure}{\arabic{figure}}
\usepackage[font=small,justification=justified,format=plain]{caption} 
\DeclareCaptionLabelSeparator{bar}{.}
\usepackage{xcolor}  % 텍스트 색상을 설정하기 위한 패키지
\usepackage[newcommands]{ragged2e} %\텍스트의 정렬 방식을 개선하고 새로운 명령어를 정의하여 텍스트 정렬을 더욱 세밀하게 제어할 수 있게 해주는 패키지
\makeatletter
\newcommand{\beginsupplement}{%
  \renewcommand{\thefigure}{S\arabic{figure}}%
  \renewcommand{\thetable}{S\arabic{table}}%
  \renewcommand{\theequation}{S\arabic{equation}}%
  \setcounter{figure}{0}%
  \setcounter{table}{0}%
  \setcounter{equation}{0}%
  \renewcommand{\ext@figure}{slos}%
}

\newcommand{\listofsuppfigures}{%
  {%
    \let\oldnumberline\numberline%
    \renewcommand{\numberline}{\figurename~\oldnumberline}%
    \section*{List of Supplementary Figures}%
    \@starttoc{slos}%
  }%
}
\makeatother
\begin{document}

\title{Imaging asymmetric Coulomb blockade phenomena across metallic nanoislands}

\author{Junho Bang}
\altaffiliation{\texorpdfstring{These authors contributed equally to this work}{}}
\affiliation{Department of Physics, Yonsei University, Seoul 03722, Republic of Korea}
\author{Byeongin Lee}
\altaffiliation{\texorpdfstring{These authors contributed equally to this work}{}}
\affiliation{Department of Physics, Yonsei University, Seoul 03722, Republic of Korea}
\author{Hankyu Lee}
\affiliation{Department of Physics, Yonsei University, Seoul 03722, Republic of Korea}
\author{Jian-Feng Ge}
\email{Jianfeng.Ge@cpfs.mpg.de}
\affiliation{Max Planck Institute for Chemical Physics of Solids, 01187 Dresden, Germany}
\author{Doohee Cho}
\email{dooheecho@yonsei.ac.kr}
\affiliation{Department of Physics, Yonsei University, Seoul 03722, Republic of Korea}

\begin{abstract}
    Coulomb blockade (CB) arises in nanoscale systems with ultra-small capacitance, where discrete charging effects dictate electron transport, enabling wide-ranging applications based on single-electron transistors. Despite established electrostatic control of charge states in quantum dots and nanoislands, a rigorous quantitative link between junction parameters and the CB spectrum remains elusive. Here, using scanning tunneling spectroscopy, we investigate the spatial variation of CB in indium nanoislands on semiconducting black phosphorus. We observe spatially dispersive charging resonances whose trajectories exhibit a finite shift of the symmetry axis in bias as well as a pronounced asymmetric curvature. By comparing the experimental results with calculations based on orthodox theory, we show that these features originate from work function differences in the junctions, underscoring the importance of junction-specific electrostatics in nanoscale charge transport.    
\end{abstract}
\maketitle

\noindent\textit{Introduction.~}The ability to detect and manipulate individual electrons has long been an essential requirement for the operation and readouts of solid-state qubits~\cite{nielsen2010quantum}, and various approaches have been developed to achieve such precise control~\cite{devoret2000amplifying, pekola2013single,shaikhaidarov2022quantized}. Among these approaches, single-electron transistors (SETs) have played a central role, for instance, serving as qubits~\cite{devoret1992single,shnirman1998quantum}, ultrasensitive charge sensors~\cite{schoelkopf1998radio,yoo1997scanning},  parity control of the superconducting ground states~\cite{vlaic2017superconducting,trivini2025local}, and precision standards in electrical metrology~\cite{flensberg1999towards, pekola2008hybrid}, thereby offering a path toward applications beyond conventional electronic devices~\cite{zwanenburg2013silicon,elzerman2004single}.

The operation of SETs is fundamentally governed by the Coulomb blockade (CB) phenomenon, which emerges from the discrete nature of electrons and electrostatic repulsion in nanoscale systems~\cite{fulton1987observation, beenakker1991theory, lafarge1993two, takahashi1998effect}. These devices can be modeled as double-barrier tunnel junctions (DBTJ), where a small metallic island is weakly coupled to the source and drain electrodes through tunnel barriers. The infinitesimal junction capacitances ($C \lesssim$ 1 aF) result in a substantial charging energy $E_\mathrm{C} \gtrsim$ 0.1 eV, which may exceed the thermal energy. At low temperatures, when an electron tunnels into the island, it prevents additional electrons from entering until the applied bias across the junction increases by an amount equal to $E_\mathrm{C}/e$, where $e$ is the charge of an electron. Overall, the tunneling spectrum of a DBTJ exhibits discrete peaks corresponding to charge-state transitions of the nanoisland~\cite{van1988incremental,deshpande2011imaging}.

Compared to planar tunnel junctions~\cite{devoret2000amplifying} and lithographically defined quantum dots~\cite{fulton1987observation}, scanning-probe-based DBTJs offer superior flexibility and precise control over tunnel junctions, allowing detailed investigations of charge dynamics and single-electron manipulation in a wide range of junction configurations~\cite{woodside2002scanned,van1988incremental,hong2013coulomb}.
In addition, their scanning capability with unprecedented spatial resolution allows further investigation of the (a)symmetry of the DBTJs. In a symmetric DBTJ with identical tunnel barriers, the electron-hole symmetry is preserved~\cite{jarillo2004electron}. In contrast, any asymmetry—arising from variations in junction geometry, barrier height, or tunneling rates—leads to unequal voltage drops under opposite bias polarities~\cite{klein1997single,negishi2007iv,wu2004control}. As a consequence, the tunneling differential conductance spectrum exhibits asymmetry with respect to zero bias, manifested as reproducible peak offsets and pronounced spatial variations of the peak energies~\cite{reiner2020spectroscopic, vlaic2017superconducting, deshpande2011imaging}. Nevertheless, existing interpretations of this spectral asymmetry, including those based on trapped charges~\cite{hanna1991variation}, polarization effects~\cite{wilkins1989scanning}, and contact potential differences~\cite{schonenberger1992single}, have not yet provided a quantitative framework that is directly related to junction properties. 

Here, we use scanning tunneling microscopy (STM) to visualize the CB phenomenon in the DBTJ formed between the STM tip and crystalline indium nanoislands on semiconducting black phosphorus (BP). Our spatially resolved tunneling spectra reveal bias-dependent dispersions of CB peaks arising from spatial variations in the junction capacitance as the STM tip scans across the nanoisland. Notably, the trajectories of the CB peaks exhibit the following asymmetries: (i) their curvatures flip sign across a bias offset from zero, and (ii) the trajectories lack mirror symmetry with respect to the offset bias. Our simulations based on orthodox theory~\cite{averin1991single} faithfully reproduce these two distinct asymmetries, and we attribute their origins to work function mismatches at the island-tip and island-substrate interfaces, respectively. These results provide a unified framework for interpreting asymmetry in nanoscale tunneling systems based on junction-specific properties.

\begin{figure}[!t]
\includegraphics{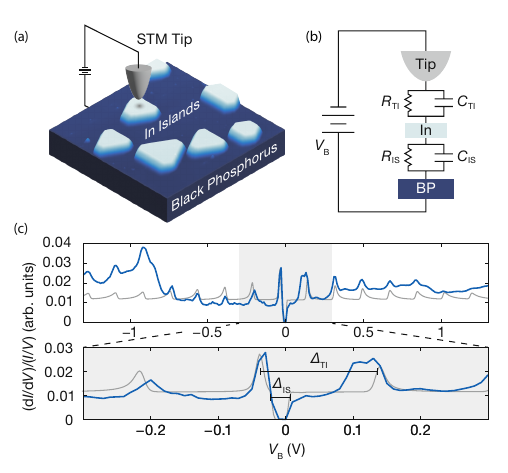}
\caption{~Coulomb blockade in an STM-defined DBTJ. (a) Schematic illustration of the STM measurement geometry, showing In nanoislands formed on a BP substrate. (b) Equivalent circuit model of the DBTJ, where $C_\mathrm{TI} \,(C_\mathrm{IS})$ and $R_\mathrm{TI} \,(R_\mathrm{IS})$ denote the capacitance and resistance between the In island and the tip (substrate), respectively. (c) Normalized differential conductance $(\mathrm{d}I/\mathrm{d}V)/(I/V)$ spectrum measured on the In nanoisland (blue), together with a simulated spectrum (gray) based on orthodox Coulomb blockade theory. Parameters of the simulation: $C_\mathrm{IS} = 6.74\,\mathrm{aF}$, $R_\mathrm{IS} = 15.0\,\mathrm{M\Omega}$, $C_\mathrm{TI} = 0.92\,\mathrm{aF}$, $R_\mathrm{TI} = 2.4\,\mathrm{G\Omega}$, and $Q_0 = 1.305\,e$. The Coulomb energy $\Delta_\mathrm{TI}$, corresponding to the peak spacing, is determined from the average separation between adjacent conductance peaks. Measurement conditions: $V_\mathrm{set} = 0.6~\mathrm{V}, I_\mathrm{set}= 250~\mathrm{pA}, V_\mathrm{mod}=10~\mathrm{mV}$ }\label{fig.1}
\end{figure}

\noindent\textit{Coulomb blockade in an STM-defined DBTJ.} Figure~\ref{fig.1}(a) shows our experimental setup to visualize the CB phenomenon using an STM tip on In nanoislands formed on BP. The nanoislands appear as truncated triangular shapes with lateral dimensions of approximately 10 nm, and one of their edges aligning with the zigzag direction of puckered honeycomb lattice of BP~\cite{SI}. All nanoislands have a uniform height of 1.3 nm, despite variations in their lateral size. Each island forms one capacitor with the STM tip and another with the underlying substrate. Owing to their nanoscale dimensions, both capacitors have extremely small capacitances ($\sim$ aF), corresponding to a large charging energy ($\sim$ 0.1 eV) far exceeding the thermal energy at 4.2 K. A DBTJ (Fig.~\ref{fig.1}(b)) is thus defined by the tunable STM junction ($R_\mathrm{TI}$ = 0.01--10 G$\Omega$) in series with the strong Schottky barrier between the metal island and semiconducting substrate~\cite{schmeidel2009coulomb,gong2014electrical}, enabling the observation of the CB phenomenon.

The measured normalized differential conductance $(\mathrm{d}I/\mathrm{d}V)/(I/V)$ spectrum (Fig.~\ref{fig.1}(c)) exhibits two characteristic features---a Coulomb gap of magnitude $e/C_\mathrm{IS}$ around zero bias and a series of conductance peaks equally spaced by $e/C_\mathrm{TI}$---in agreement with previous studies~\cite{hong2013coulomb, reiner2020spectroscopic}. These features are well captured by the orthodox theory of correlated electron tunneling in an asymmetric DBTJ~\cite{averin1991single,hanna1991variation}. The Coulomb gap arises from the electrostatic energy required to add an extra electron to the island, leading to charge quantization governed by the island-substrate capacitance $C_\mathrm{IS}$. This capacitance scales with nanoisland area $A$ as $C_\mathrm{IS} = \epsilon_0\epsilon_r A / d_\mathrm{eff}$, where $d_\mathrm{eff}$ is the effective island-substrate separation, $\epsilon_0$ the vacuum permittivity, and $\epsilon_r$ the relative dielectric constant~\cite{SI}. This periodic charging becomes more pronounced when the tunneling resistances of the two junctions are compatible ($R_\mathrm{TI} \approx R_\mathrm{IS}$)~\cite{SI}. In contrast, in the limit $R_\mathrm{TI} \gg R_\mathrm{IS}$, electron transport is dominated by the tunneling junction between the tip and nanoisland, and the conductance spectrum is governed by charging effects associated with $C_\mathrm{TI}$. The experimentally observed spectrum is well reproduced by a simulation based on the orthodox theory (gray curve in Fig.~\ref{fig.1}(c)) using the extracted values of $C_\mathrm{TI}$ and $C_\mathrm{IS}$ which further allows determination of the corresponding junction resistances. 

\begin{figure}[b]
\includegraphics{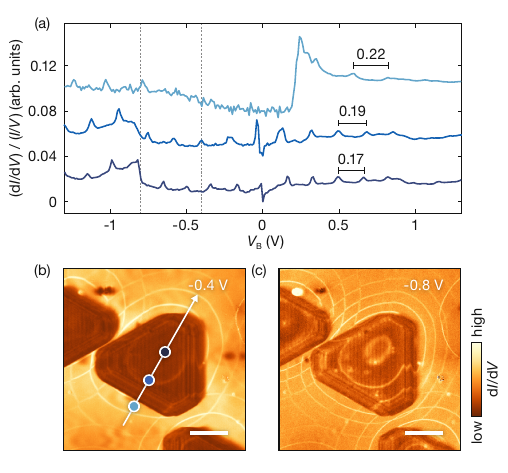}
\caption{~Spatially resolved Coulomb blockade spectra. (a), Normalized differential conductance $(\mathrm{d}I/\mathrm{d}V)/(I/V)$ spectra acquired at different lateral positions across an In nanoisland, as indicated by the colored markers in (b). The horizontal bars denote the extracted Coulomb peak spacings $\Delta_\mathrm{TI}$. Setup conditions: $V_\mathrm{set} = 0.6~\mathrm{V}, I_\mathrm{set}= 250~\mathrm{pA}, V_\mathrm{mod}=10~\mathrm{mV}$. (b) and (c)  Differential conductance maps of an island at different bias voltages, showing concentric conductance modulations around the nanoisland. Scale bar, $10~\mathrm{nm}$. Setup conditions: $V_\mathrm{set}=  -0.4 ~{\rm V}\ \mathrm{and} -0.8~{\rm V}$ respectively,  $I_\mathrm{set}= 100~\mathrm{pA}, V_\mathrm{mod}=10~\mathrm{mV}$. }\label{fig.2}
\end{figure}

\noindent\textit{Spatially resolved Coulomb blockade.} To elucidate the spatial dependence of the CB phenomenon, we investigate how the charging characteristics evolve as the STM tip is laterally displaced across the nanoisland. With the ability to precisely control the lateral position of the STM tip, we can continuously tune the effective capacitive coupling in the DBTJ. Figure~\ref{fig.2}(a) compares tunneling spectra acquired at various lateral positions on the island, revealing a systematic increase in the CB peak spacing $\Delta_\mathrm{TI}$ as the tip moves away from the island center. Notably, CB peaks remain clearly observable even when the tip is positioned outside the physical boundary of the nanoisland (light blue curve), accompanied by a further increase in $\Delta_\mathrm{TI}$. These observations indicate that the capacitive coupling between the tip and the island is governed by the overall junction geometry~\cite{SI} and extends beyond the physical extent of the island, even in the absence of direct resistive tunneling. In this way, the STM-defined DBTJ functions as a reconfigurable single-electron box, in which the charging energy can be continuously tuned via the lateral displacement of the tip, enabling nonlocal control of the island charge states~\cite{kouwenhoven1997introduction,reiner2020spectroscopic}.

The spatial variation of the CB peaks is also indicated in the $\mathrm{d}I/\mathrm{d}V$ maps (Fig.~\ref{fig.2}(b)–(c)), which are acquired in the same field of view but at different bias voltages, marked by dashed lines in Fig.~\ref{fig.2}(a). Of particular interest is the emergence of concentric conductance rings centered around the In islands, exhibiting a slight triangular distortion. Along these rings, CB peaks appear at a certain bias, whereas at positions between the rings the peaks are shifted out of the measurement bias voltage, as evidenced by the spectra in Fig.~\ref{fig.2}(a). The presence of multiple concentric rings demonstrates that the charge state of the island is discretely tuned by repositioning the tip at a fixed bias voltage. Regions between adjacent rings correspond to spatial domains in which the island maintains the same charge state. Moreover, as the absolute value of the bias increases~\cite{SI}, these rings extend outward from the island center and become more densely spaced, indicating a nonuniform spatial dispersion of CB peaks. Such spatially resolved charging features, widely reported in scanning-probe studies of quantum dots~\cite{pradhan2005atomic,woodside2002scanned,cockins2010energy}, demonstrate that the charge state of individual dots can be remotely manipulated with single-electron precision.

\noindent\textit{Bias asymmetry in spatially dispersive Coulomb peaks.} The main focus of this work is the asymmetries appearing in spatially resolved CB spectra. The asymmetric behavior is implied in the spatial variation of the spectra and the conductance maps, such that the CB peaks show distinct energy values and nonuniform spatial separation at opposite bias polarities. Despite the phenomenological understanding in previous quantum-dot experiments, a quantitative connection between these asymmetries and the physical properties of the tunneling junctions has not been thoroughly elucidated thus far. 

\begin{figure}[b]
\includegraphics{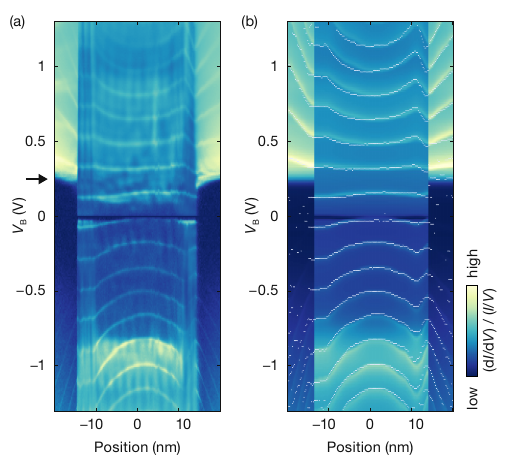}
\caption{~Spatially dispersive Coulomb peaks in an In nanoisland. (a) Spatially resolved normalized $(\mathrm{d}I/\mathrm{d}V)/(I/V)$ spectra acquired across an In nanoisland. The curved features correspond to CB peaks whose energies vary with lateral position. Setup conditions: $V_\mathrm{set} = 0.6~\mathrm{V}, I_\mathrm{set}= 250~\mathrm{pA}, V_\mathrm{mod}=10~\mathrm{mV}$. (b) Orthodox theory-based simulation with %work function, $\phi_\mathrm{I(In)}\simeq4.12~\mathrm{eV}$~\cite{michaelson1977work}, $\phi_\mathrm{S(BP)} = 4.5\simeq\mathrm{eV}$ for 5 layers~\cite{cai2014layer}, and 
the work-function differences, $\delta\phi_\mathrm{TI} = 0.25~\mathrm{eV}$ and $\delta\phi_\mathrm{IS} = -0.3784~\mathrm{eV}$, chosen within the range of the reference values to achieve the best agreement with the experimental data~\cite{SI}.  The background signals of in- and outside the island are from averaged data of each region in (a).  The white dots mark the CB peak positions extracted from the measured spectra in (a). Parameters of simulation: $C_\mathrm{IS} = 6.74\,\mathrm{aF}$, $R_\mathrm{IS} = 15.0\,\mathrm{M\Omega}$, $C_\mathrm{TI} = 0.8\text{--}1.0\,\mathrm{aF}$, $R_\mathrm{TI} = 2.4\,\mathrm{G\Omega}$. }\label{fig.3}
\end{figure}

To directly visualize the asymmetries, we performed spatially resolved $\mathrm{d}I/\mathrm{d}V$ measurements to track the CB peaks in energy and in real space along the white arrow indicated in Fig.~\ref{fig.2}(b). The resulting line spectra (Fig.~\ref{fig.3}(a)) reveal multiple CB peak trajectories that appear as vertically offset curves whose curvature increases with the absolute value of the bias voltage. The CB peak spacing varies continuously across the island, directly reflecting the spatial variations of $C_\mathrm{TI}$. Specifically, as the STM tip moves away from the island center, the effective $C_\mathrm{TI}$ decreases~\cite{SI}. This spatial gradient of $C_\mathrm{TI}$ causes cumulative shifts in the CB peak positions, giving rise to progressively larger shifts of the CB peaks with increasing positive (or decreasing negative) bias. This result explains the tendency of the concentric $\mathrm{d}I/\mathrm{d}V$ rings to appear more closely packed as the absolute value of the bias increases.

However, the spatially varying capacitive coupling between the tip and the island cannot explain the pronounced asymmetry with respect to the bias polarity. The line $\mathrm{d}I/\mathrm{d}V$ spectra exhibit a clear asymmetry in the curvature of the CB peak trajectories: the sign of the curvature does not reverse as the bias voltage reverses sign at $V_{\rm B}=0$, but instead near a finite bias offset of $+0.25\ \mathrm{V}$ (black arrow in Fig.~\ref{fig.3}(a)). Even when the trajectories are compared relative to this offset, those appearing at biases below the offset exhibit a larger curvature than those above it. This behavior causes the more sharpened lines and more tightly spaced concentric patterns observed at negative bias compared to positive bias.

We emphasize that such asymmetric CB behavior cannot be resolved through either point spectroscopy or spectroscopic imaging at a fixed bias. While point spectra do show a shift in CB peaks, the measurable range of this shift is limited to less than half of the CB gap because the peaks appear periodically with a spacing equal to the gap~\cite{hong2013coulomb}. On the other hand, constant-bias images do show the spatial variation of the CB peaks, their bias dependence---and more importantly the asymmetry in bias polarity---are overlooked. In contrast, the spatially resolved spectra capture the full energy dispersion of the CB peaks, revealing the deviations from the expected symmetry with detail that allows quantitative analysis. As shown in Fig.~\ref{fig.3}(b), our simulations not only recover both asymmetries present in our experimental data, but also are in excellent, quantitative agreement with all the CB peak energies as a function of spatial position (white dots) extracted from Fig.~\ref{fig.3}(a).

\begin{figure}[b]
\includegraphics{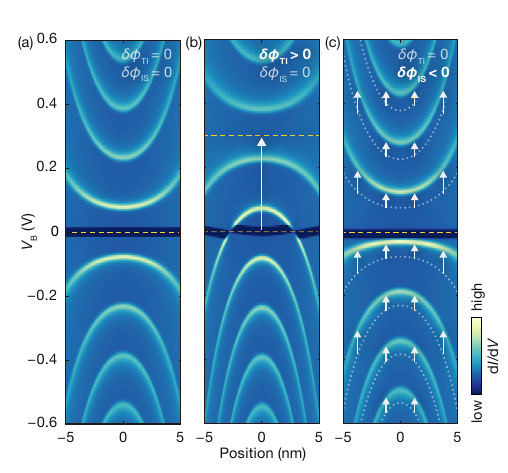}
\caption{~Coulomb blockade asymmetry from work function differences. (a)--(c) Simulated $\mathrm{d}I/\mathrm{d}V$ spectra as a function of $C_\mathrm{TI}$ with varying work function differences: (a), no work function difference, (b), between tip and island ($\delta\phi_\mathrm{TI} = 0.3\,\mathrm{eV}$) and (c), between island and substrate ($\delta\phi_\mathrm{IS} = -0.007\,\mathrm{eV}$). The white arrows in (b) and (c) show the amount of the shift due to the $\delta \phi _\mathrm{TI}$ and $\delta\phi_\mathrm{IS}$ respectively, compared to the (a) (no work function difference). The yellow dotted lines in (a--c) represent the dispersion axis. In (b), the axis is shifted by amount of $\delta \phi _\mathrm{TI}$, compared to the axis of (a) marked as shaded dotted line. The white dotted curves in (c) mark the CB curvatures of (a). Parameters of simulation: $C_\mathrm{IS} = 6.74\,\mathrm{aF}$, $R_\mathrm{IS} = 15.0\,\mathrm{M\Omega}$, $C_\mathrm{TI} = 4.5\text{--}10.5\,\mathrm{aF}$, $R_\mathrm{TI} = 2.4\,\mathrm{G\Omega}$.}\label{fig.4}
\end{figure}

\noindent\textit{Origin of asymmetric Coulomb blockade phenomena.} To understand our observations, we begin with the orthodox theory of Coulomb blockade and invoke the language of the residual charge $ Q_0 $. This residual charge arises from work function mismatch in the two interfaces above and below the nanoisland~\cite{hanna1991variation}
\[
Q_0 = (C_\mathrm{TI} \delta\phi_\mathrm{TI} - C_\mathrm{IS} \delta\phi_\mathrm{IS})/e \ ,
\] 
where the work function differences are defined as $ \delta\phi_\mathrm{TI} = \phi_\mathrm{T} -\phi_\mathrm{I}$ and  $ \delta\phi_\mathrm{IS} = \phi_\mathrm{I} -\phi_\mathrm{S} $, with $\phi_{T}$, $\phi_{I}$, and $\phi_{S}$ being the work functions of the tip, the island, and the substrate, respectively. The residual charge has previously been attributed to CB peak shifts~\cite{hong2013coulomb}. While capacitance variations only rescale the energy spacing between CB peaks with respect to a fixed reference point (e.g., zero bias for an ideal symmetric DBTJ), the residual charge shifts all spectral features. In the simplest case of symmetric DBTJs where the components have identical work functions, the residual charge becomes zero, and the dispersion should appear symmetric around zero bias, as shown in Fig.~\ref{fig.4}(a). However, since work function depends on material and geometric details, having two electrodes with identical work functions is not realistic and residual charge exists inevitably. Consequently, most DBTJs, including our system, in practice exhibit a shift in their CB peaks. 

In this framework, we can analyze the distinct influence of the two work function differences on the bias and spatial dependence of CB peaks. For simplicity and better illustration, we assume a spatial variation in $C_\mathrm{TI}$ that is quadratic with position and reaches its maximum at the center of the island, similar to our experimental setup. Then we simulate the spectrum by varying one of $ \delta\phi_\mathrm{TI} $ and $ \delta\phi_\mathrm{IS} $ and fixing the other, and vice versa. 

We first examine the case of variation of $\delta\phi_\mathrm{TI}$, by comparing it with the reference case of no work function difference (Fig.~\ref{fig.4}(a)). The reference case shows the spatial dispersion of CB peaks symmetric about the zero bias axis. However, a finite tip–island work function mismatch ($\delta\phi_\mathrm{TI} > 0$) shows a rigid shift in bias of the entire dispersion structure as indicated by the white arrow in Fig.~\ref{fig.4}(b), while maintaining the same trajectory profile with no curvature changes. The displacement is equal to $\delta\phi_\mathrm{TI}$ itself and is independent of the local $ C_\mathrm{TI} $ variation. In this case, the residual charge can be quantitatively determined by finding the axis of the spatial dispersion of the CB peaks. Notably, the residual charge can exceed a single electron charge $e$ , whereas in single-point spectroscopy only a fractional electron charge can be inferred due to the periodic nature of CB spectral features. Next, we vary $ \delta\phi_\mathrm{IS} $ while keeping $ \delta\phi_\mathrm{TI} = 0 $, and examine the variation of the peak energies and positions. Unlike $\delta\phi_\mathrm{TI}$ variations which cause a rigid shift in bias without curvature changes, the CB trajectories undergo both energy shifts and curvature changes (white arrows in Fig.~\ref{fig.4}(c)): the displacement of peaks varies with the local capacitance $C_\mathrm{TI}$; the curvatures of individual trajectories are modified accordingly, exhibiting an asymmetry about zero bias.  

Based on the above discussion, we can quantitatively dissect our data and simulation in Fig.~\ref{fig.3}. First, from the axis where curvature reverses sign occurring at $V_\mathrm{B} = +~0.25$ V, we directly obtain the work function mismatch between the tip and island $\delta\phi_\mathrm{TI} = 0.25$ eV. Second, with extracted spatial variation of $C_\mathrm{TI}$ by the average $\Delta_\mathrm{TI}$ in each point spectrum, we find the  work function mismatch $\delta\phi_\mathrm{IS} = -0.378$ eV. Along with the resistance $R_\mathrm{IS}=15.0~\mathrm{M\Omega}$ estimated from $R_\mathrm{TI}$-dependent spectra~\cite{SI}, and capacitance $C_\mathrm{IS} = 6.74$ aF obtained from point spectra (Fig.~\ref{fig.1}), we therefore experimentally obtain the full set of junction parameters ($R_i, C_i, \delta\phi_i,~ i=\mathrm{TI, IS}$) by matching the simulated CB peak trajectories  with the observed ones. Regarding the asymmetries, we attribute both the nonzero curvature-sign-reversal bias and the asymmetric curvatures to the division of $C_\mathrm{TI}$ in measuring real observable voltage $V_0 = eQ_0/C_\mathrm{TI}$, as the STM junction defines the measurable quantities. Therefore, all the CB peaks reflect themselves in a voltage shift determined by  $eV_0 = \delta \phi_\mathrm{TI}-C_\mathrm{IS}\delta\phi_\mathrm{IS}/C_\mathrm{TI}$, where the first part $\delta \phi_\mathrm{TI}$ is a constant while the second part depends on the spatial profile of $C_\mathrm{TI}$.
We note, however, that variation of $\delta\phi_\mathrm{IS}$ may lead to a periodic change of the resulting CB trajectories~\cite{SI}. Care then must be taken when choosing the size of the island to acquire the junction parameters, as a larger island has a larger $C_\mathrm{IS}$ and thus a smaller periodicity ($e^2/C_\mathrm{IS}$) such that multiple values of $\delta\phi_\mathrm{IS}$ can simulate the same observed spectra. The above procedures are thus applicable to quantum-dot-based single-electron devices, whose junction parameters can be quantitatively obtained by our spatially resolved Coulomb-blockade spectroscopy.

\noindent\textit{Conclusion.} In conclusion, we have demonstrated a spatially resolved analysis of the CB phenomenon in a DBTJ formed between an STM tip and In nanoislands on BP. By mapping the curvature and symmetry of CB peak dispersion, we disentangled the distinct electrostatic roles of the tip-island and island-substrate work function differences. The tip-island work function mismatch ($\delta\phi_\mathrm{TI}$) determines the mirroring (offset) voltage in CB peak dispersion, while the island-substrate work function mismatch ($\delta\phi_\mathrm{IS}$) governs the asymmetry in peak curvature. These effects are quantitatively captured by simulations based on orthodox theory, enabling the extraction of key junction parameters from spectral features alone. These findings establish spectroscopic-imaging Coulomb blockade microscopy~\cite{qian2010imaging} as a protocol for quantitative diagnosis of junction parameters. Furthermore, the persistence of CB peaks outside the physical island area indicates long-range capacitive coupling, suggesting the feasibility of remote single-electron control via lateral electrostatic environment in nanoscale tunnel junctions, offering critical insights into precision charge-state control at the nanoscale in single-electron transistor devices. 

\bibliographystyle{apsrev4-2-title} %Physical Review 
\bibliography{bib_CB}

\newpage
%%%% Figure 1 %%%%
\captionsetup{justification=Justified}

\newpage
%%%% Figure 2 %%%%
\captionsetup{justification=Justified}

\newpage
%%%% Figure 3 %%%%
\captionsetup{justification=Justified}

\newpage
%%%% Figure 4 %%%%
\captionsetup{justification=Justified}

\clearpage

% ---- SI title block ----
\begin{center}
{\large \textmd{Supplementary Material for} \newline Imaging asymmetric Coulomb blockade phenomena across metallic nanoislands}

\end{center}
\thispagestyle{empty}
% ------------------------

\beginsupplement

\listofsuppfigures

\bigskip
\bigskip
\noindent{\bf Methods: Sample preparation and STM measurement}

\noindent\textit{Sample preparation.} Single crystals of BP (HQ Graphene) were cleaved at room temperature in an ultra-high vacuum (UHV) environment. Indium was thermally evaporated onto the BP substrate at a cell temperature of $723\,\mathrm{K}$ for $6$ min. The sample was kept at room temperature during the deposition. The coverage of In atoms is roughly $0.67\,\mathrm{ML}$.

\noindent\textit{STM and STS measurements.} STM and STS measurements were performed using a commercial low-temperature STM (UNISOKU-USM1200) at $4.2\,\mathrm{K}$ in the UHV environment down to $1\times10^{-10}\,\mathrm{Torr}$.The mechanically sharpened Pt-Ir (90/10) wires were used for STM tips and characterized on Au(111) single crystal before measurements. STM images were obtained in constant current mode by applying bias voltages $V_\mathrm{set}$ between the sample and the tip. The differential conductance ($\mathrm{d}I/\mathrm{d}V$) images were simultaneously acquired with the STM images using a standard lock-in technique, an a.c. voltage modulation with an amplitude $V_\mathrm{mod} = 10\,\mathrm{mV}$ and frequency $f_\mathrm{mod} = 613\,\mathrm{Hz}$ added to the d.c. sample bias.

\newpage
\captionsetup{justification=Justified}
\begin{figure}[htb!]
\includegraphics[scale=1]{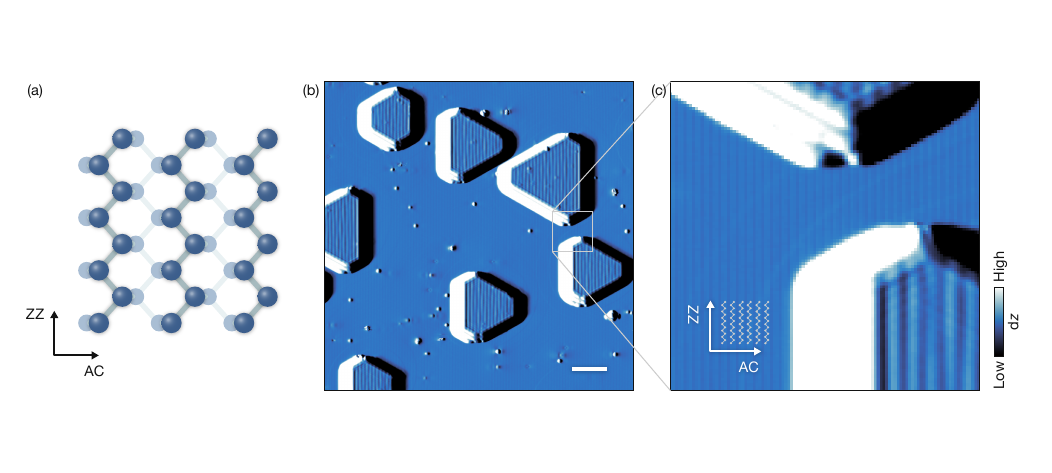}
\caption[Atomic structures of In islands and BP substrate]{~Atomic structures of In islands and BP substrate. (a) Schematic representation of black phosphorus (BP) atomic structure showing zigzag-shaped chains puckered out-of-plane. The zigzag (ZZ) and armchair (AC) directions are indicated. (b) Differential topography of In islands on BP, showing various islands, adatoms which do not form islands, and vacancy defects of BP. Triangular islands align one edge with the ZZ direction. Scale bar, $10~\mathrm{nm}$. (c) Magnified view of (b) clearly showing the BP lattice orientation and island alignment. Stripes on islands are moir\'e patterns resulting from In-BP lattice mismatch.}\label{Fig.S1} \end{figure}

\newpage
\captionsetup{justification=Justified}
\begin{figure}[htb!]
\includegraphics[scale=1]{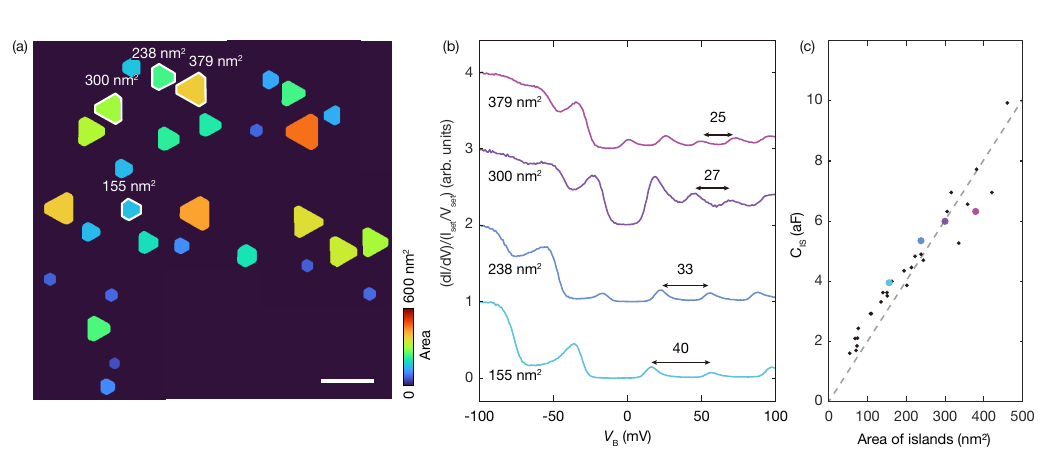}
\caption[Size-dependent Coulomb gap ($\Delta_\mathrm{IS}=e/C_\mathrm{IS}$) in In nanoislands.]{~Size-dependent Coulomb gap ($\Delta_\mathrm{IS}=e/C_\mathrm{IS}$) in In nanoislands. (a) Filtered topography of In islands, with each island color-coded according to its area ($A$). Scale bar, $2~\mathrm{nm}$.
(b) Normalized $\mathrm{d}I/\mathrm{d}V$ spectra acquired at the centers of islands with different areas. The charging energy decreases with increasing island size. Each curve is equally shifted for clarity. 
(c) Scatter plot of $C_{\rm IS}$ versus island area for 30 individual islands. Colored points correspond to the spectra shown in b. The measured capacitances agree well with the linear fit (dashed gray line), $C_{\rm IS} = (\epsilon_{\rm eff}/d_{\rm eff})A$, where $d_{\rm eff}$ is the effective separation between the island and the substrate, and $\epsilon_{\rm eff}$ is the corresponding effective permittivity.}\label{Fig.S2}\end{figure}

\newpage
\captionsetup{justification=Justified}
\begin{figure}[htb!]
\includegraphics[scale=1]{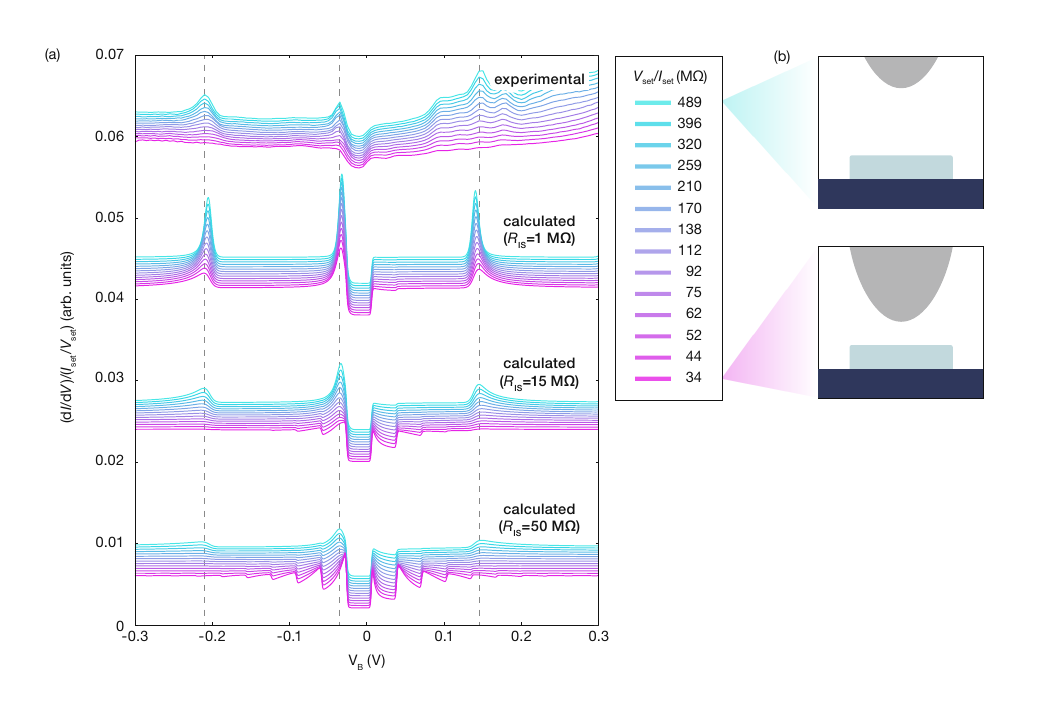}
\caption[Evolution of CB spectral features with varying junction resistances.]{~Evolution of Coulomb-blockade spectral features with varying junction resistances. (a) $\mathrm{d}I/\mathrm{d}V$ spectra normalized by the total conductance ($I_\mathrm{set}/V_\mathrm{set}$) while tuning the tip–island distance via the setpoint current $I_\mathrm{set}$ at a fixed $V_\mathrm{set}=+0.6~\mathrm{V}$. The corresponding tunnel resistances $V_\mathrm{set}/I_\mathrm{set}$ span from $34$ to $489~\mathrm{M\Omega}$, and these values are used as $R_\mathrm{TI}$ in the simulations. The top traces show the experimental spectra, while the three traces below show spectra calculated within the orthodox theory for $R_\mathrm{IS}=1,15,50~\mathrm{M\Omega}$. All spectra are vertically offset for clarity. Reducing $R_\mathrm{TI}$ results in broadened and suppressed Coulomb peaks (dashed lines), accompanied by the emergence of oscillatory features whose periodicity corresponds to the Coulomb gap associated with $e/C_\mathrm{IS}$ rather than $e/C_\mathrm{TI}$. At $R_\mathrm{IS}=1~\mathrm{M\Omega}$, Coulomb peaks persist across the entire $R_\mathrm{TI}$ range,  while oscillatory features remain suppressed at low $R_\mathrm{TI}$. In contrast, at $R_\mathrm{IS}=50~\mathrm{M\Omega}$, Coulomb peaks are almost suppressed even at high $R_\mathrm{TI}$ and the oscillatory features become pronounced. Considering the evolutions of both Coulomb peaks and oscillatory features, we choose $R_\mathrm{IS}=15~\mathrm{M\Omega}$ as the simulation parameter.
(b) Schematic illustration of the STM tip–island junction geometry for high (cyan) and low (magenta) tunnel-resistance configurations.}
\label{Fig.S3}
\end{figure}

\newpage
\captionsetup{justification=Justified}
\begin{figure}[htb!]
\includegraphics[scale=1]{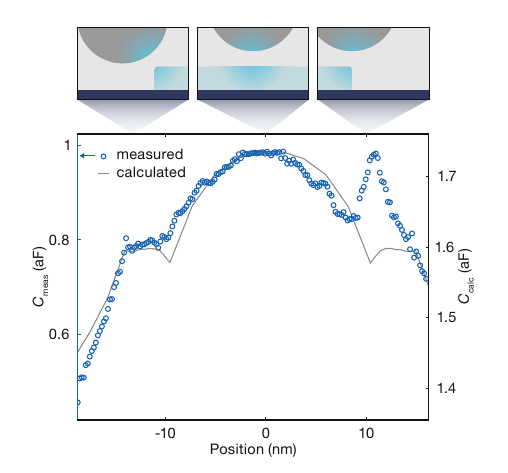}
\caption[Numerical capacitance calculation using \textsc{comsol}]{~Numerical capacitance calculation using \textsc{comsol}. Blue circles indicate the capacitance values extracted from $\Delta_\mathrm{TI}$ measurements, while the black solid line shows the corresponding numerically calculated capacitance $C_\mathrm{TI}$. The schematics above illustrate the tip–island junction geometry when the tip is positioned at the center and at the edge of the island.}
\label{Fig.S4}
\end{figure}

\newpage
\captionsetup{justification=Justified}
\begin{figure}[htb!]
\includegraphics[scale=1]{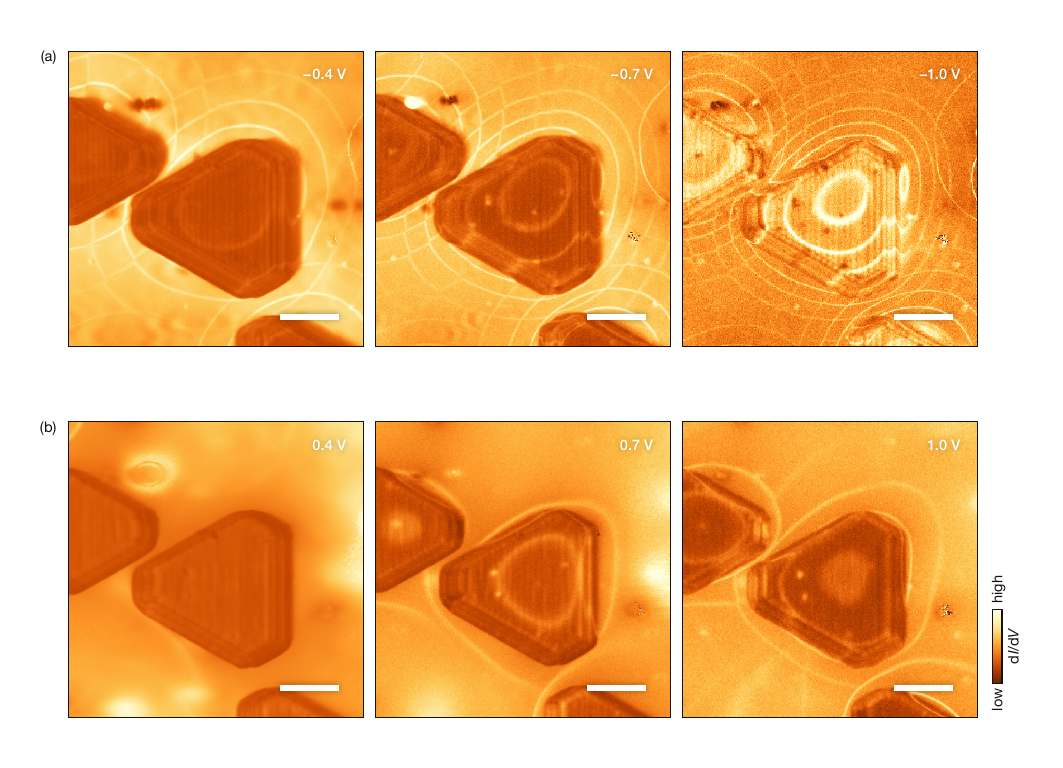}
\caption[Bias dependency of CB ring patterns]{~Bias dependency of CB ring patterns. (a) Differential conductance maps of the same spatial frame at negative bias voltages of $-0.4$, $-0.7$, $-1.0$~V, and (b) at positive bias voltages of $0.4$, $0.7$, $1.0$~V, respectively. For both polarities with increasing bias voltage magnitude, the rings become clearer with narrower linewidth and a reduced inter-ring distance. Scale bar, 10 nm.}
\label{Fig.S5}
\end{figure}

\newpage
\captionsetup{justification=Justified}
\begin{figure}[htb!]
\includegraphics[scale=1]{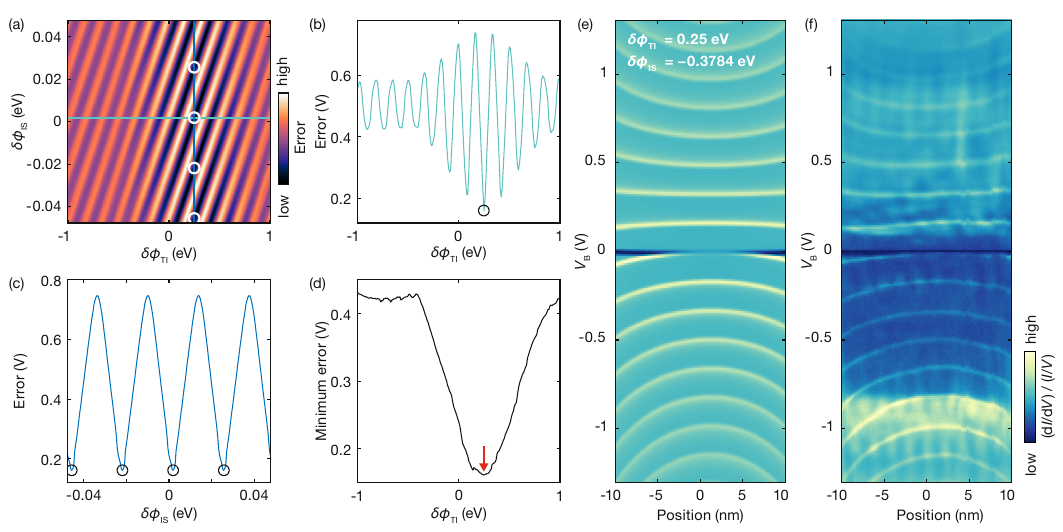}
\caption[Determination of optimal $\delta\phi_\mathrm{TI}$ and $\delta\phi_\mathrm{IS}$.]{~Determination of optimal $\delta\phi_\mathrm{TI}$ and $\delta\phi_\mathrm{IS}$. (a) Two-dimensional error map showing the deviation between experimental and simulated CB peaks as a function of $\delta\phi_\mathrm{TI}$ and $\delta\phi_\mathrm{IS}$. The error is defined as the root mean square (RMS) of the voltage differences between experimental and simulated CB peaks. White circles mark the optimal parameter sets. (b) Line cut of the error map along the $\delta\phi_\mathrm{TI}$ direction at the optimal $\delta\phi_\mathrm{IS}$ (mint line in (a)). (c) Line cut along the $\delta\phi_\mathrm{IS}$ direction at the optimal $\delta\phi_\mathrm{TI}$ (blue line in (a)). Errors on $\delta\phi_\mathrm{IS}$ axis show periodic behavior, which reflects the periodic nature of $Q_0$. Black circles mark the values for $\delta\phi_\mathrm{IS}$ with identical, minimum error. (d) Minimum error as a function of $\delta\phi_\mathrm{TI}$, obtained by minimizing over $\delta\phi_\mathrm{IS}$ at each $\delta\phi_\mathrm{TI}$ value. The red arrow indicates the position of error-minimizing $\delta\phi_\mathrm{TI}$. The parameters for the minimum error are $\delta\phi_\mathrm{TI}^* = 0.25~\mathrm{eV}$, and $\delta\phi_\mathrm{IS}^* = 0.0019~\mathrm{eV} +(e^2/C_\mathrm{IS})n$, where $n$ is an integer. (e) Simulated spectra using the optimal parameter $\delta\phi_\mathrm{TI}^*$ and $\delta\phi_\mathrm{IS}^* = -0.3784~\mathrm{eV}$. The value of $\delta\phi_\mathrm{IS}^*$ is a reasonable estimate, given by $\phi(\mathrm{In})-\phi(\mathrm{BP)} \approx -0.38~\mathrm{eV}$ \cite{}. (f) Experimental spectra acquired on the nanoisland cropped from Fig. 3. For error estimation, only the peak positions of spectra on the island are used.}\label{Fig.S6}
\end{figure}

\newpage
\captionsetup{justification=Justified}
\begin{figure}[htb!]
\includegraphics[scale=1]{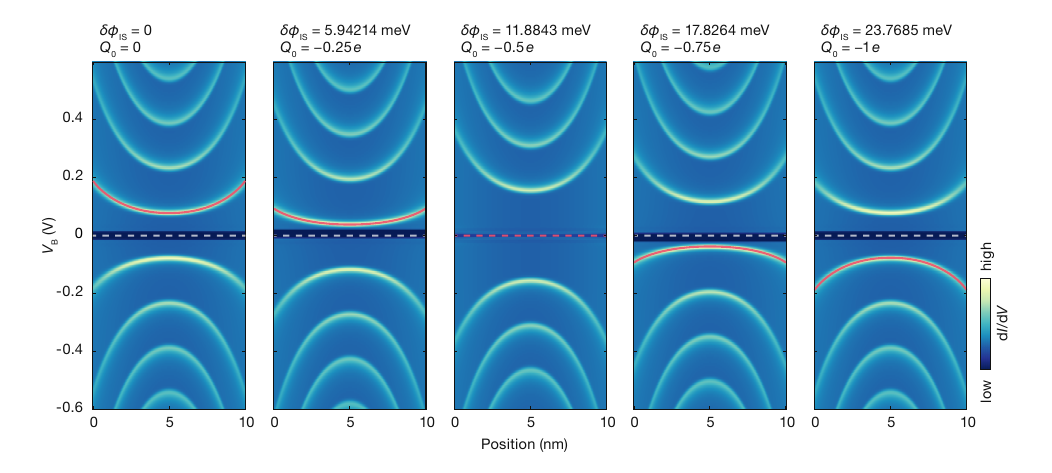}
\caption[Evolution of the dispersion curve with varying $\mathbf{\delta\phi_\mathrm{IS}}$.]{~Evolution of the dispersion curve with varying $\mathbf{\delta\phi_\mathrm{IS}}$. The spatial dispersion of the CB peaks evolves systematically as $\delta\phi_\mathrm{IS}$ increases, corresponding to residual charges $Q_0 = 0,-0.25e,-0.5e,-0.75e,$ and $-1e$. The dashed line marks the dispersion axis defined by variations in $C_\mathrm{TI}$. The dispersion curves for all $\delta\phi_\mathrm{IS}$ values trace an identical trajectory along this axis. Following the convex curve nearest to the Fermi level (indicated by the red line), the curvature gradually flattens toward $Q_0=-0.5e$, and becomes concave once $Q_0$ exceeds $-0.5e$. When  $Q_0$ reaches $-1e$, the curve evolves to become symmetric to its shape at $Q_0=0$ with respect to the axis. Overall, the dispersion exhibits a periodic evolution in energy, with a periodicity of $e$ in residual charge $Q_0$, or equivalently a periodicity of $e^2/C_\mathrm{IS}$ in $\delta\phi_\mathrm{IS}$, because when $\delta\phi_\mathrm{TI} =0$, $Q_0 \propto C_\mathrm{IS}\delta\phi_\mathrm{IS}/e$. In practice, for an unambiguous determination of $\delta\phi_\mathrm{IS}$, the periodicity $e^2/C_\mathrm{IS}$ should exceed the typical uncertainty of work function ($\sim$ 0.1 eV). Simulation parameters: $C_\mathrm{IS}=4.70~\mathrm{aF}$, $R_\mathrm{IS}=15.0~\mathrm{M\Omega}$, $C_\mathrm{TI}=0.49\text{–}0.59~\mathrm{aF}$, $R_\mathrm{TI}=2.4~\mathrm{G\Omega}$, $\delta\phi_\mathrm{TI}=0$.}\label{Fig.S7}
\end{figure}

\end{document}